\documentclass[12pt,thmsa]{article}
\def \eq {\begin{equation}}
\def \fim-eq {\end{equation}}
\begin{document}

\author{E. S. Guerra \\
%EndAName
Departamento de F\'{\i}sica \\
Universidade Federal Rural do Rio de Janeiro \\
Cx. Postal 23851, 23890-000 Serop\'edica, RJ, Brazil \\
email:emerson@ufrrj.br}
\title{TELEPORTATION OF THE STATES OF MOTION OF ATOMS BY INTERACTION WITH
TWO-SLIT SCREENS \ AND CAVITIES}
\maketitle

\begin{abstract}
\noindent We present a scheme of teleportation of the state of motion of
atoms making use of screens with slits and cavities. \ The fascinating
aspects of quantum mechanics are highlighted making use of entanglement and
the interaciton of atoms with slits.

\ \newline



%\date{The Date }
\end{abstract}

\section{Introduction}

\ Only about twenty years ago it was put forward by Bonioff \cite{Benioff}
and Feynmann \cite{Feynmann} the idea of building up of computers based on
the principles of quantum mechanics which was developed concretely by David
Deutsch \cite{Deutsch}\ in 1985. Since then it has been discussed in the
literature the possible implications of quantum entanglement \cite%
{Schrodinger} and non-locality in the broad new and promising filed of
research embracing quantum information and quantum computation \cite%
{Nielsen, MathQC, PQI}. Entanglement and non-locality is fascinating, and
although at the same time it is conterintuitive, it can have revolutionizing
impact on our thinking about information processing and computing. Einstein
Podolsky and Rosen, noticed the implications of quantum entanglement and
proposed a \textit{gedanken }experiment, the EPR experiment \cite{EPR} in
order to show that quantum mechanics was not a complete theory which could
explain nature correctly. Probably one of the most notable and dramatic
among various concepts developed through the application of quantum
mechanics to the information science and which is a consequence of
entanglement and non-locality, is teleportation put forward by Bennett 
\textit{et al} \cite{Bennett}, given rise to a new field of research.
Quantum teleportation is an experimental reality and it holds tremendous
potential for applications in the fields of quantum communication and
computing \cite{MathQC, Nielsen, PQI}. For instance, it can be used \ to
build quantum gates which are resistant to noise and is intimately connected
with quantum error-correcting codes \cite{Nielsen}.

As Feymann put it, "the double-slit experiment has in it the heart of
quantum mechanics". The double-slit experiment has been performed with many
different kinds of particles ranging from photons \cite{DLphotons},
electrons \cite{DLelectrons}, neutrons \cite{DLneutrons} and atoms \cite%
{DLatoms}. In this article we propose a scheme of teleportation of the
states of motion of atoms interacting with two-slit screens and the
electromagnetic field in cavities. We assume that the atoms are Rydberg
atoms with relatively long radiative lifetime \cite{Rydat}. We also assume
that the cavities are microwave superconducting cavities of relatively large
quality factors \cite{haroche, walther}

\section{TELEPORTATION\ OF\ THE\ STATE\ OF\ NOTION\ OF\ ATOMS}

Let us assume two three-level lambda atoms $A1$ prepared in the state $\
\mid b_{1}\rangle $ and $A2$ prepared in the state $\ \mid b_{2}\rangle $ \
to pass through a double slit screen $SC1$ where behind slit $SL1$ (at $%
\zeta _{1})$ there is a cavity $C1$ prepared in a coherent state $|\alpha
_{1}\rangle $ and behind slit $SL2$ \ (at $\zeta _{2})$ there is a cavity $%
C2 $ prepared in a coherent state $|\alpha _{2}\rangle $ \cite{Louisell,
Orszag}$.$ Now, for a three-level lambda atom interacting with the
electromagnetic field inside a cavity where the \ upper and the two
degenerated lower states are $|a\rangle ,$ $|b\rangle $ and $|c\rangle $
respectively, and for which the $|a\rangle \rightleftharpoons |c\rangle $
and $|a\rangle \rightleftharpoons |b\rangle $ transitions are in the far
from resonance interaction limit, the time evolution operator $U(t)$ for the
atom-field interaction in a cavity $Ck$ is given by \cite{Knight}%
\begin{eqnarray}
U(\tau ) &=&-e^{i\varphi a_{k}^{\dagger }a_{k}}|a\rangle \langle a|+\frac{1}{%
2}(e^{i\varphi a_{k}^{\dagger }a_{k}}+1)|b\rangle \langle b|+\frac{1}{2}%
(e^{i\varphi a_{k}^{\dagger }a_{k}}-1)|b\rangle \langle c|\ +  \nonumber \\
&&\frac{1}{2}(e^{i\varphi a_{k}^{\dagger }a_{k}}-1)|c\rangle \langle b|+%
\frac{1}{2}(e^{i\varphi a_{k}^{\dagger }a_{k}}+1)|c\rangle \langle c|,
\end{eqnarray}%
where $a_{k}$ $(a_{k}^{\dagger })$ is the annihilation (creation) operator
for the field in cavity $Ck$, $\varphi =2g^{2}\tau /$ $\Delta $, \ $g$ is
the coupling constant, $\Delta =\omega _{a}-\omega _{b}-\omega =\omega
_{a}-\omega _{c}-\omega $ is the detuning where \ $\omega _{a}$, $\omega
_{b} $ and $\omega _{c}$\ are the frequencies of the upper and \ of the two
degenerate lower levels respectively and $\omega $ is the cavity field
frequency and $\tau $ is the atom-field interaction time. For $\varphi =\pi $%
, we get 
\begin{equation}
U(\tau )=-\exp \left( i\pi a_{k}^{\dagger }a_{k}\right) |a\rangle \langle
a|+\Pi _{k,+}|b\rangle \langle b|+\Pi _{k,-}|b\rangle \langle c|\ +\Pi
_{k,-}|c\rangle \langle b|+\Pi _{k,+}|c\rangle \langle c|,  \label{UlambdaPi}
\end{equation}%
where 
\begin{eqnarray}
\Pi _{k,+} &=&\frac{1}{2}(e^{i\pi a_{k}^{\dagger }a_{k}}+1),  \nonumber \\
\Pi _{k,-} &=&\frac{1}{2}(e^{i\pi a_{k}^{\dagger }a_{k}}-1),  \label{pi+-}
\end{eqnarray}%
and we have 
\begin{eqnarray}
\Pi _{k,+}|+\rangle _{k} &=&|+\rangle _{k},  \nonumber \\
\Pi _{k,+}|-\rangle _{k} &=&0,  \nonumber \\
\Pi _{k,-}|-\rangle _{k} &=&-|-\rangle _{k},  \nonumber \\
\Pi _{k,-}|+\rangle _{k} &=&0,
\end{eqnarray}%
where the even coherent state $|+\rangle _{k}$ and odd coherent state $%
|-\rangle _{k}$ are given by 
\begin{equation}
\mid \pm \rangle _{k}=\mid \alpha _{k}\rangle \pm \mid -\alpha _{k}\rangle
\label{EOCS}
\end{equation}%
\cite{EvenOddCS}, and we have used $e^{za_{k}^{\dagger }a_{k}}|\alpha
_{k}\rangle =|e^{z}\alpha _{k}\rangle $ \cite{Louisell}.

Taking into account (\ref{UlambdaPi}), after atom $A1$ has passed through
the slits of screen $SC1$ and before it has passed through the cavities we
have%
\begin{equation}
|\psi \rangle _{A1}=\frac{1}{\sqrt{2}}(|\zeta _{11}\rangle +|\zeta
_{12}\rangle )|\alpha _{1}\rangle |\alpha _{2}\rangle \mid b_{1}\rangle .
\end{equation}%
After the atom $A1$ passes through the cavities we get%
\begin{equation}
|\psi \rangle _{A1}=\frac{1}{2\sqrt{2}}\{|\zeta _{11}\rangle (|b_{1}\rangle
|+\rangle _{1}-|c_{1}\rangle |-\rangle _{1})|\alpha _{2}\rangle +|\zeta
_{12}\rangle (|b_{1}\rangle |+\rangle _{2}-|c_{1}\rangle |-\rangle
_{2})|\alpha _{1}\rangle \}.
\end{equation}%
Then, after atom $\ A2$ passes through the two slits $SL1$ and $SL2$ \ and
before it has interacted with the cavities we have%
\begin{eqnarray}
|\psi \rangle _{A1-A2} &=&\frac{1}{4}\{|\zeta _{21}\rangle +|\zeta
_{22}\rangle \}\{|\zeta _{11}\rangle (|b_{1}\rangle |+\rangle
_{1}-|c_{1}\rangle |-\rangle _{1})|\alpha _{2}\rangle + \\
&&|\zeta _{12}\rangle (|b_{1}\rangle |+\rangle _{2}-|c_{1}\rangle |-\rangle
_{2})|\alpha _{1}\rangle \}|b_{2}\rangle ,
\end{eqnarray}%
and after $A2$ has passed through $C1$ and $C2$ we have%
\begin{eqnarray}
|\psi \rangle _{A1-A2} &=&\frac{1}{4}\{|\zeta _{21}\rangle |\zeta
_{11}\rangle (|+\rangle _{1}|b_{1}\rangle |b_{2}\rangle +|-\rangle
_{1}|c_{1}\rangle |c_{2}\rangle )|\alpha _{2}\rangle  \nonumber \\
&&+|\zeta _{21}\rangle |\zeta _{12}\rangle (|+\rangle _{2}|b_{1}\rangle
-|-\rangle _{2}|c_{1}\rangle )\frac{1}{2}|+\rangle _{1}|b_{2}\rangle 
\nonumber \\
&&+|\zeta _{22}\rangle |\zeta _{11}\rangle (|+\rangle _{1}|b_{1}\rangle
-|-\rangle _{1}|c_{1}\rangle )\frac{1}{2}|+\rangle _{2}|b_{2}\rangle + 
\nonumber \\
&&+|\zeta _{22}\rangle |\zeta _{12}\rangle \}(|b_{1}\rangle |b_{2}\rangle
|+\rangle _{2}+|c_{1}\rangle |c_{2}\rangle |-\rangle _{2})|\alpha
_{1}\rangle \}.
\end{eqnarray}%
Now if we detect $|c_{1}\rangle $ and $|b_{2}\rangle $ we get%
\begin{equation}
|\psi \rangle _{A1-A2}=\frac{1}{N}\{|\zeta _{21}\rangle |\zeta _{12}\rangle
|-\rangle _{2}|+\rangle _{1}+|\zeta _{22}\rangle |\zeta _{11}\rangle
|-\rangle _{1}|+\rangle _{2}\}.
\end{equation}%
where $N$ is a normalization constant. Now we let an atom $A3$ prepared in
the state%
\begin{equation}
\bigskip |\psi \rangle _{A3}=\frac{1}{\sqrt{2}}(c_{b}|b_{3}\rangle
-c_{c}|c_{3}\rangle ).
\end{equation}%
to fly through the slits and cavities of $SC1$. After $A3$ has passed
through the slits we have%
\begin{equation}
|\psi \rangle _{A1-A2-A3}=\frac{1}{N}\{|\zeta _{31}\rangle +|\zeta
_{32}\rangle \}\{|\zeta _{21}\rangle |\zeta _{12}\rangle |-\rangle
_{2}|+\rangle _{1}+|\zeta _{22}\rangle |\zeta _{11}\rangle |-\rangle
_{1}|+\rangle _{2}\}(c_{b}|b_{3}\rangle -c_{c}|c_{3}\rangle ),
\end{equation}%
and after $A3$ has passed through the cavities we have%
\begin{eqnarray}
|\psi \rangle _{A1-A2-A3} &=&\frac{1}{N}\{|\zeta _{31}\rangle |\zeta
_{21}\rangle |\zeta _{12}\rangle |-\rangle _{2}|+\rangle
_{1}c_{b}|b_{3}\rangle +  \nonumber \\
&&-|\zeta _{31}\rangle |\zeta _{21}\rangle |\zeta _{12}\rangle |-\rangle
_{2}|+\rangle _{1}c_{c}|c_{3}\rangle +  \nonumber \\
&&-|\zeta _{31}\rangle |\zeta _{22}\rangle |\zeta _{11}\rangle |-\rangle
_{1}|+\rangle _{2}c_{b}|c_{3}\rangle +  \nonumber \\
&&|\zeta _{31}\rangle |\zeta _{22}\rangle |\zeta _{11}\rangle |-\rangle
_{1}|+\rangle _{2}c_{c}|b_{3}\rangle +  \nonumber \\
&&-|\zeta _{32}\rangle |\zeta _{21}\rangle |\zeta _{12}\rangle |-\rangle
_{2}|+\rangle _{1}c_{b}|c_{3}\rangle +  \nonumber \\
&&|\zeta _{32}\rangle |\zeta _{21}\rangle |\zeta _{12}\rangle |-\rangle
_{2}|+\rangle _{1}c_{c}|b_{3}\rangle +  \nonumber \\
&&|\zeta _{32}\rangle |\zeta _{22}\rangle |\zeta _{11}\rangle |-\rangle
_{1}|+\rangle _{2}c_{b}|b_{3}\rangle +  \nonumber \\
&&-|\zeta _{32}\rangle |\zeta _{22}\rangle |\zeta _{11}\rangle |-\rangle
_{1}|+\rangle _{2}c_{c}|c_{3}\rangle \}.
\end{eqnarray}%
If we detect $|b_{3}\rangle $ we have%
\begin{eqnarray}
|\psi \rangle _{A1-A2-A3} &=&\frac{1}{N}\{|\zeta _{31}\rangle |\zeta
_{21}\rangle |\zeta _{12}\rangle |-\rangle _{2}|+\rangle _{1}c_{b}|+ 
\nonumber \\
&&|\zeta _{31}\rangle |\zeta _{22}\rangle |\zeta _{11}\rangle |-\rangle
_{1}|+\rangle _{2}c_{c}+  \nonumber \\
&&|\zeta _{32}\rangle |\zeta _{21}\rangle |\zeta _{12}\rangle |-\rangle
_{2}|+\rangle _{1}c_{c}+  \nonumber \\
&&|\zeta _{32}\rangle |\zeta _{22}\rangle |\zeta _{11}\rangle |-\rangle
_{1}|+\rangle _{2}c_{b}\}.
\end{eqnarray}

Now, we let atom $A3$ evolve toward a screen $SC2$ near screen $SC1$ with
one slit and a detector behind $|\zeta \rangle $ and detect $A3$ at $|\zeta
\rangle $. If the time evolution operator for the evolution of $A3$ toward
the screen $SC2$ is $U(t,0)$ we have, writing $\langle \zeta |U(t,0)$ $%
|\zeta _{31}\rangle =\psi _{\zeta _{31}}(\zeta )$ and $\langle \zeta |U(t,0)$
$|\zeta _{32}\rangle =\psi _{\zeta _{32}}(\zeta )$, 
\begin{eqnarray}
|\psi \rangle _{A1-A2-A3} &=&\frac{1}{N}\{\psi _{\zeta _{31}}(\zeta )|\zeta
\rangle |\zeta _{21}\rangle |\zeta _{12}\rangle |-\rangle _{2}|+\rangle
_{1}c_{b}+  \nonumber \\
&&\psi _{\zeta _{31}}(\zeta )|\zeta \rangle |\zeta _{22}\rangle |\zeta
_{11}\rangle |-\rangle _{1}|+\rangle _{2}c_{c}+  \nonumber \\
&&\psi _{\zeta _{32}}(\zeta )|\zeta \rangle |\zeta _{21}\rangle |\zeta
_{12}\rangle |-\rangle _{2}|+\rangle _{1}c_{c}+  \nonumber \\
&&\psi _{\zeta _{32}}(\zeta )|\zeta \rangle |\zeta _{22}\rangle |\zeta
_{11}\rangle |-\rangle _{1}|+\rangle _{2}c_{b}\}.
\end{eqnarray}%
For $|\zeta \rangle =|\zeta _{31}\rangle $ and setting $\psi _{\zeta
_{32}}(\zeta _{31})=0$ since we assume that $SC1$ and $SC2$ are near enough
such that this condition be true, we get%
\begin{equation}
|\psi \rangle _{A1-A2-A3}=\frac{1}{N}\{c_{b}|\zeta _{21}\rangle |\zeta
_{12}\rangle |-\rangle _{2}|+\rangle _{1}+c_{c}|\zeta _{22}\rangle |\zeta
_{11}\rangle |-\rangle _{1}|+\rangle _{2}\},
\end{equation}%
which are the states of motion of atoms $A1$ and $A2$ entangled with the
cavity field states. Now we let atom $A1$ to pass through a double slit
screen $SC3$ with $SL3$ at $|\gamma _{1}\rangle $ and $SL4$ at $|\gamma
_{2}\rangle $. Before $A1$ passes through these slits, making use of $%
|\gamma _{1}\rangle \langle \gamma _{1}|+|\gamma _{2}\rangle \langle \gamma
_{2}|=1$, we write%
\begin{equation}
|\psi \rangle _{A1-A2}=\frac{1}{N}\{|\gamma _{1}\rangle \langle \gamma
_{1}|+|\gamma _{2}\rangle \langle \gamma _{2}|\}\{c_{b}|\zeta _{21}\rangle
|\zeta _{12}\rangle |-\rangle _{2}|+\rangle _{1}+c_{c}|\zeta _{22}\rangle
|\zeta _{11}\rangle |-\rangle _{1}|+\rangle _{2}\},
\end{equation}%
and writing for the evolution toward $SC3$ the operator $U(t,0)$ and $%
\langle \gamma _{1}|U(t,0)$ $|\zeta _{12}\rangle =\psi _{\zeta _{21}}(\gamma
_{1}),$ $\langle \gamma _{1}|U(t,0)$ $|\zeta _{11}\rangle =\psi _{\zeta
_{11}}(\gamma _{1}),\langle \gamma _{2}|U(t,0)$ $|\zeta _{12}\rangle =\psi
_{\zeta _{12}}(\gamma _{2})$ and $\langle \gamma _{2}|U(t,0)$ $|\zeta
_{11}\rangle =\psi _{\zeta _{11}}(\gamma _{2})$ we have%
\begin{eqnarray}
|\psi \rangle _{A1-A2} &=&\frac{1}{N}\{c_{b}\psi _{\zeta _{12}}(\gamma
_{1})|\gamma _{1}\rangle |\zeta _{21}\rangle |-\rangle _{2}|+\rangle _{1}+ 
\nonumber \\
&&c_{c}\psi _{\zeta _{11}}(\gamma _{1})|\gamma _{1}\rangle |\zeta
_{22}\rangle |-\rangle _{1}|+\rangle _{2}+  \nonumber \\
&&c_{b}\psi _{\zeta _{12}}(\gamma _{2})|\gamma _{2}\rangle |\zeta
_{21}\rangle |-\rangle _{2}|+\rangle _{1}+  \nonumber \\
&&c_{c}\psi _{\zeta _{11}}(\gamma _{2})|\gamma _{2}\rangle |\zeta
_{22}\rangle |-\rangle _{1}|+\rangle _{2}\}.
\end{eqnarray}%
If we detect atom $A1$ using a screen $SC4$ with one slit at $|\zeta \rangle
=|\gamma _{1}\rangle $ and a detector behind this slit, we get%
\begin{equation}
|\psi \rangle _{A2}=\frac{1}{N}\{c_{b}\psi _{\zeta _{12}}(\gamma _{1})|\zeta
_{21}\rangle |-\rangle _{2}|+\rangle _{1}+c_{c}\psi _{\zeta _{11}}(\gamma
_{1})|\zeta _{22}\rangle |-\rangle _{1}|+\rangle _{2}\}.
\end{equation}%
Now, we assume that atom $A1$ is far enough from the screen $SC3$ so that
the spherical waves $\psi _{\zeta _{12}}(\gamma _{1})$ and $\psi _{\zeta
_{11}}(\gamma _{1})$ \ have evolved in such a way that we can consider $\psi
_{\zeta _{12}}(\gamma _{1})\cong $ $\psi _{\zeta _{11}}(\gamma _{1})$ $\ $%
and therefore we can write%
\begin{equation}
|\psi \rangle _{A2}=\frac{1}{N}\{c_{b}|\zeta _{21}\rangle |-\rangle
_{2}|+\rangle _{1}+c_{c}|\zeta _{22}\rangle |-\rangle _{1}|+\rangle _{2}\}.
\label{Telepst1}
\end{equation}%
This is the state of motion of $A2$ which is far apart from the screen $SC1$
to be teleported to another atom $A4$. In order to do so we pass an atom $A4$
through $SL1$ and $Sl2$ in screen $SC1$ and $C1$ and $C2$. Then we have,
after $A4$ has passed through the slits%
\begin{equation}
|\psi \rangle _{A2-A4}=\frac{1}{N}\{|\zeta _{41}\rangle +|\zeta _{42}\rangle
\}\{c_{b}|\zeta _{21}\rangle |-\rangle _{2}|+\rangle _{1}+c_{c}|\zeta
_{22}\rangle |-\rangle _{1}|+\rangle _{2}\}|b_{4}\rangle ,
\end{equation}%
and after $A4$ has passed through the cavities%
\begin{eqnarray}
|\psi \rangle _{A2-A4} &=&\frac{1}{N}\{c_{b}|\zeta _{41}\rangle |\zeta
_{21}\rangle |-\rangle _{2}|+\rangle _{1}|b_{4}\rangle -c_{c}|\zeta
_{41}\rangle |\zeta _{22}\rangle |-\rangle _{1}|+\rangle _{2}|c_{4}\rangle 
\nonumber \\
&&-c_{b}|\zeta _{42}\rangle |\zeta _{21}\rangle |-\rangle _{2}|+\rangle
_{1}|c_{4}\rangle +c_{c}|\zeta _{42}\rangle |\zeta _{22}\rangle |-\rangle
_{1}|+\rangle _{2}|b_{4}\rangle \}.
\end{eqnarray}%
Now, we let atom $A2$ to pass through a screen $SC5$ with two slits $SL5$
(at $\rho _{1}$) and $SL6$ (at $\rho _{2}$) and before $A2$ passes through
the slits, using $|\rho _{1}\rangle \langle \rho _{1}|+|\rho _{2}\rangle
\langle \rho _{2}|=1$, we have 
\begin{eqnarray}
|\psi \rangle _{A2-A4} &=&\frac{1}{N}\{|\rho _{1}\rangle \langle \rho
_{1}|+|\rho _{2}\rangle \langle \rho _{2}|\}\{c_{b}|\zeta _{41}\rangle
|\zeta _{21}\rangle |-\rangle _{2}|+\rangle _{1}|b_{4}\rangle -c_{c}|\zeta
_{41}\rangle |\zeta _{22}\rangle |-\rangle _{1}|+\rangle _{2}|c_{4}\rangle 
\nonumber \\
&&-c_{b}|\zeta _{42}\rangle |\zeta _{21}\rangle |-\rangle _{2}|+\rangle
_{1}|c_{4}\rangle +c_{c}|\zeta _{42}\rangle |\zeta _{22}\rangle |-\rangle
_{1}|+\rangle _{2}|b_{4}\rangle \},
\end{eqnarray}%
and writing $\langle \rho _{1}|U(t,0)$ $|\zeta _{21}\rangle =\psi _{\zeta
_{21}}(\rho _{1})$, $\langle \rho _{1}|U(t,0)$ $|\zeta _{22}\rangle =\psi
_{\zeta _{42}}(\rho _{1})$, $\langle \rho _{2}|U(t,0)$ $|\zeta _{21}\rangle
=\psi _{\zeta _{21}}(\rho _{2})$ and $\langle \rho _{2}|U(t,0)$ $|\zeta
_{22}\rangle =\psi _{\zeta _{22}}(\rho _{2})$ we have%
\begin{eqnarray}
|\psi \rangle _{A2-A4} &=&\frac{1}{N}\{c_{b}\psi _{\zeta _{21}}(\rho
_{1})|\rho _{1}\rangle |\zeta _{41}\rangle |-\rangle _{2}|+\rangle
_{1}|b_{4}\rangle +  \nonumber \\
&&-c_{c}\psi _{\zeta _{22}}(\rho _{1})|\rho _{1}\rangle |\zeta _{41}\rangle
|-\rangle _{1}|+\rangle _{2}|c_{4}\rangle +  \nonumber \\
&&-c_{b}\psi _{\zeta _{21}}(\rho 1)|\rho _{1}\rangle |\zeta _{42}\rangle
|-\rangle _{2}|+\rangle _{1}|c_{4}\rangle +  \nonumber \\
&&c_{c}\psi _{\zeta _{22}}(\rho _{1})|\rho _{1}\rangle |\zeta _{42}\rangle
|-\rangle _{1}|+\rangle _{2}|b_{4}\rangle +  \nonumber \\
&&c_{b}\psi _{\zeta _{21}}(\rho _{2})|\rho _{2}\rangle |\zeta _{41}\rangle
|-\rangle _{1}|+\rangle _{2}|b_{4}\rangle +  \nonumber \\
&&-c_{c}\psi _{\zeta _{22}}(\rho _{2})|\rho _{2}\rangle |\zeta _{41}\rangle
|-\rangle _{1}|+\rangle _{2}|c_{4}\rangle +  \nonumber \\
&&-c_{b}\psi _{\zeta _{21}}(\rho _{2})|\rho _{2}\rangle |\zeta _{42}\rangle
|-\rangle _{2}|and+\rangle _{1}|c_{4}\rangle +  \nonumber \\
&&c_{c}\psi _{\zeta _{22}}(\rho _{2})|\rho _{2}\rangle |\zeta _{42}\rangle
|-\rangle _{1}|+\rangle _{2}|b_{4}\rangle \}.
\end{eqnarray}%
If we detect atom $A2$ using a screen $SC6$ with one slit at $|\zeta \rangle
=$ $|\rho _{1}\rangle $ and a detector behind this slit, we get%
\begin{eqnarray}
|\psi \rangle _{A2-A4} &=&\frac{1}{N}\{c_{b}\psi _{\zeta _{21}}(\rho
_{1})|\zeta _{41}\rangle |-\rangle _{2}|+\rangle _{1}|b_{4}\rangle + 
\nonumber \\
&&-c_{c}\psi _{\zeta _{22}}(\rho _{1})|\zeta _{41}\rangle |-\rangle
_{1}|+\rangle _{2}|c_{4}\rangle +  \nonumber \\
&&-c_{b}\psi _{\zeta _{21}}(\rho 1)|\zeta _{42}\rangle |-\rangle
_{2}|+\rangle _{1}|c_{4}\rangle +  \nonumber \\
&&c_{c}\psi _{\zeta _{22}}(\rho _{1})|\zeta _{42}\rangle |-\rangle
_{1}|+\rangle _{2}|b_{4}\rangle \},.  \nonumber
\end{eqnarray}%
and then we detect $|b_{4}\rangle $ and we get%
\begin{equation}
|\psi \rangle _{A4}=\frac{1}{N}\{c_{b}\psi _{\zeta _{21}}(\rho _{1})|\zeta
_{41}\rangle |-\rangle _{2}|+\rangle _{1}+c_{c}\psi _{\zeta _{22}}(\rho
_{1})|\zeta _{42}\rangle |-\rangle _{1}|+\rangle _{2}\}.  \nonumber
\end{equation}%
Now, we assume that atom $A2$ is far enough from the screen $SC5$ so that
the spherical waves $\psi _{\zeta _{21}}(\rho _{1})$ and $\psi _{\zeta
_{22}}(\rho _{1})$\ have evolved in such a way that we can consider $\psi
_{\zeta _{21}}(\rho _{1})\cong $ $\psi _{\zeta _{22}}(\rho _{1})\ $and
therefore we can write%
\begin{equation}
|\psi \rangle _{A4}=\frac{1}{N}\{c_{b}|\zeta _{41}\rangle |-\rangle
_{2}|+\rangle _{1}+c_{c}|\zeta _{42}\rangle |-\rangle _{1}|+\rangle _{2}\}.
\label{Telepst2}
\end{equation}%
and we have teleported the state of motion of $A2$ (\ref{Telepst1}) to the
state of motion of $A4$ (\ref{Telepst2}).We can now disentangle that state
of motion of $A4$ from the cavity fields states. If we inject $|\alpha
_{1}\rangle $ and $|\alpha _{2}\rangle $ in cavities $C1$ and $C2$
respectively we have%
\begin{equation}
|\psi \rangle _{A4}=\frac{1}{N}\{c_{b}|\zeta _{41}\rangle (\mid 2\alpha
_{2}\rangle -\mid 0_{2}\rangle )(\mid 2\alpha _{1}\rangle +\mid 0_{1}\rangle
)+c_{c}|\zeta _{42}\rangle (\mid 2\alpha _{1}\rangle -\mid 0_{1}\rangle
)(\mid 2\alpha _{2}\rangle +\mid 0_{2}\rangle )\}.
\end{equation}

In order to disentangle the atomic states of the cavity field state we now
send two two-level atoms $A5k$ ($k=1,2)$, resonant with the cavity, with $%
|f_{5k}\rangle $ and $|e_{5k}\rangle $ being the lower and upper levels
respectively, through $Ck$. If $A5k$ is sent in the lower state $%
|f_{5k}\rangle $, under the Jaynes-Cummings dynamics \cite{Orszag} we know
that the state $|f_{5k}\rangle |0\rangle $ does not evolve, however, the
state $|f_{5k}\rangle |2\alpha _{k}\rangle $ evolves to $|e_{5k}\rangle
|\chi _{e5k}\rangle +|f_{5k}\rangle |\chi _{f5k}\rangle $, where $|\chi
_{f5k}\rangle =\sum\limits_{n_{k}}C_{n_{k}}\cos (gt\sqrt{n})|n_{k}\rangle $
and $|\chi _{e5k}\rangle =-i\sum\limits_{n_{k}}C_{n_{k}}\sin (gt\sqrt{n+1}%
)|n_{k}\rangle $ and $C_{n_{k}}=e^{-\frac{1}{2}|2\alpha _{k}|^{2}}(2\alpha
_{k})^{n_{k}}/\sqrt{n_{k}!}$. Then we get%
\begin{eqnarray}
|\psi \rangle _{A4-A3} &=&  \nonumber \\
\frac{1}{N}\{c_{b}|\zeta _{41}\rangle \lbrack |e_{52}\rangle |\chi
_{e52}\rangle +|f_{52}\rangle |\chi _{f52}\rangle -|f_{52}\rangle &\mid
&0_{2}\rangle ][|e_{51}\rangle |\chi _{e51}\rangle +|f_{51}\rangle |\chi
_{f51}\rangle +|f_{351}\rangle \mid 0_{1}\rangle ]+  \nonumber \\
c_{c}|\zeta _{42}\rangle \lbrack |e_{51}\rangle |\chi _{e51}\rangle
+|f_{51}\rangle |\chi _{f51}\rangle -|f_{51}\rangle &\mid &0_{1}\rangle
][|e_{52}\rangle |\chi _{e52}\rangle +|f_{52}\rangle |\chi _{f52}\rangle
+|f_{52}\rangle \mid 0_{2}\rangle ]\},
\end{eqnarray}%
and if we detect atom $A5k$ in state $|e_{5k}\rangle $ finally we get 
\begin{equation}
|\psi \rangle _{A4}=\frac{1}{N}\{c_{b}|\zeta _{41}\rangle +c_{c}|\zeta
_{42}\rangle \},
\end{equation}%
and the teleportation is completed successfully. In the above
disentanglement process we can choose coherent field states with a
photon-number distribution with a sharp peak at average photon number $%
\langle n_{k}\rangle =|\alpha _{k}|^{2}$ so that, to a good approximation, $%
|\chi _{f5k}\rangle \cong C_{\langle n_{k}\rangle }\cos (\sqrt{\langle
n_{k}\rangle }g\tau )|\overline{n_{k}}\rangle $ and $|\chi _{e5k}\rangle
\cong C_{\langle n_{k}\rangle }\sin (\sqrt{\langle n_{k}\rangle }g\tau )|%
\overline{n_{k}}\rangle $, where $\overline{n_{k}}$ is the integer nearest $%
\langle n_{k}\rangle $, and we could choose, for instance $\ \sqrt{\langle
n_{k}\rangle }g\tau =\pi /2$, so that we would have $|\chi _{e5k}\rangle
\cong C_{\langle n\rangle }|\overline{n_{k}}\rangle $ and $|\chi
_{f5k}\rangle \cong 0$. In this case atom $A5k$ \ would be detected in state 
$|e_{5k}\rangle $ with almost $100\%$ of probability. Therefore, proceeding
this way, we can guarantee that the atomic and field states will be
disentangled successfully as we would like. In Fig. 1 we show the set-up of
the teleportation experiment.

Concluding, we have presented a scheme of teleportation of the spacial state
(a state of the continuum) of atoms making use of cavities and screens with
slits. \ Teleportation shows dramatically the strange and fascinating
behavior we can find in quantum mechanics due to entanglement. The
experiments involving screens with slits also shows the strange and
fascinating aspects of quantum mechanics. In this work we have combined
teleportation with slit experiments and shown in one strike the beautiful
and fascinating aspects of quantum mechanics.

\bigskip

\textbf{Figure Caption}

\bigskip

\textbf{Fig. 1 - }Experimental set-up of the teleportation. Atom $A1$ pass
through the screen $SC1$, cavities $C1$ and $C2$, detector $D1$, screens $SC3
$ and $SC4$ and finally its position is detected in detector $DP1$. Atom $A2$
pass through the screen $SC1$, cavities $C1$ and $C2$, detector $D2$,
screens $SC5$ and $SC6$ and finally its position is detected in detector $DP2
$. Atom $A3$ pass through the screen $SC1$, cavities $C1$ and $C2$, detector 
$D3$, screens $SC2$ and finally its position is detected in detector $DP3$.
\ Atom $A4$ pass through the screen $SC1$, cavities $C1$ and $C2$, \ and
detector $D4$. Atoms $A51$ and $A52$ are not shown in the figure.

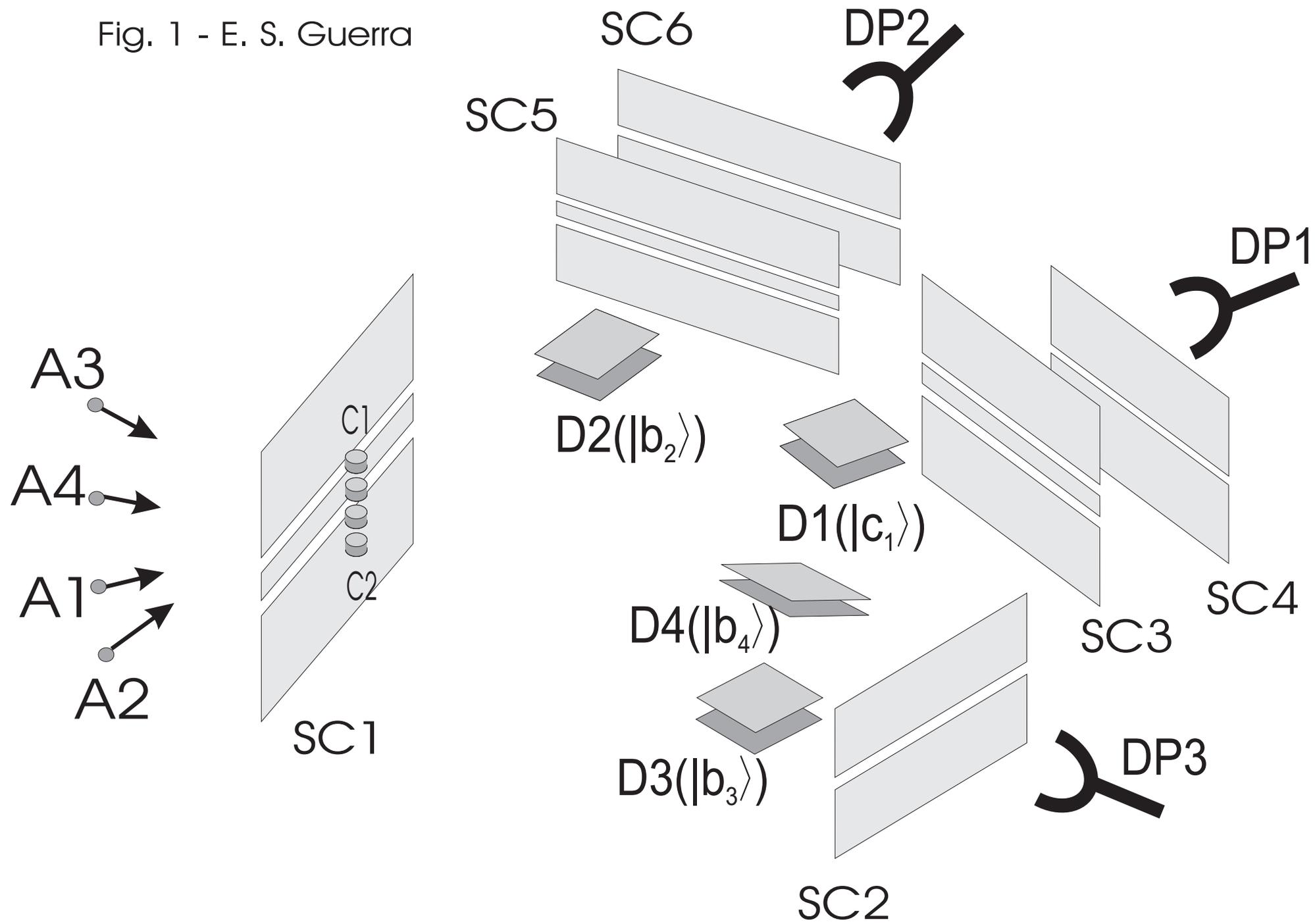
\end{document}